\documentclass[%
 reprint,
 amsmath,amssymb,
 aps,showpacs,pra,letterpaper, superscriptaddress]{revtex4-1}
\usepackage{braket}
\usepackage{graphicx}
\usepackage{dcolumn}
\usepackage{bm}
\usepackage{epstopdf}
\usepackage{comment}
\usepackage{float}
\newcommand{\vc}[1]{\mathbf{#1}}

\begin{document}


\title{Floquet Heating in Interacting Atomic Gases with an Oscillating Force}

\author{Jun-Ru Li}
\thanks{Present address: JILA, Department of Physics, University of Colorado, 440 UCB, Boulder, Colorado 80309, USA.}
\thanks{junru.li@colorado.edu}
\affiliation{Research Laboratory of Electronics, MIT-Harvard Center for Ultracold Atoms, Department of Physics,
Massachusetts Institute of Technology, Cambridge, Massachusetts 02139, USA}
\author{Boris Shteynas}
\affiliation{Research Laboratory of Electronics, MIT-Harvard Center for Ultracold Atoms, Department of Physics,
Massachusetts Institute of Technology, Cambridge, Massachusetts 02139, USA}
\author{Wolfgang Ketterle}
\affiliation{Research Laboratory of Electronics, MIT-Harvard Center for Ultracold Atoms, Department of Physics,
Massachusetts Institute of Technology, Cambridge, Massachusetts 02139, USA}

\date{\today}

\begin{abstract}
We theoretically investigate the collisional heating of a cold atom system subjected to time-periodic forces. We show within the Floquet framework that this heating rate due to two-body collisions has a general semiclassical expression $\mathcal{P}\propto \rho \sigma v_{\rm col} E_0$, depending on the kinetic energy $E_0$ associated with the shaking, particle number density $\rho$, elastic collision cross section $\sigma$,  and an effective collisional velocity $v_{\rm col}$ determined by the dominant energy scale in the system. We further show that the collisional heating is suppressed by Pauli blocking in cold fermionic systems, and by the modified density of states in systems in lower dimensions. Our results provide an exactly solvable example and reveal some general features of Floquet heating in interacting systems.
\end{abstract}


\maketitle


\section{Introduction}

Engineering novel Hamiltonians is central to quantum simulations. In general, Hamiltonians can be implemented directly and statically, or in a time-averaged way. The latter implies periodic driving of the system. If the fast modulation can be neglected, an effective time-averaged static Hamiltonian is realized as formally captured by \textit{Floquet theory}~\cite{Rahav2003,Goldman2014}. With proper driving, dynamically generated \textit{Floquet Hamiltonians} can be designed. Such \textit{Floquet systems} potentially exhibit novel properties which are difficult or impossible to be realized in static settings. Examples include synthetic gauge fields~\cite{Hauke2012, Struck2014,Creffield2016,Bukov2016}, spin-orbit coupling~\cite{Anderson2013,Xu2013, Jimenez2015}, 
and topological bands and materials~\cite{Yu2016,Baur2014, Cayssol2013,Lindner2013}. Experimental progress includes creation of the Hofstadter-Hamiltonian in optical lattices for neutral atoms~\cite{Aidelsburger2013,Miyake2013,Kennedy2015,Aidelsburger2014}, realization of the topological Haldane model with shaken optical lattice~\cite{Jotzu2014}, and the demonstration of dressed recoil momentum for radio-frequency photons in ultracold gases with modulated magnetic fields~\cite{Shteynas2018}. 

However, \textit{higher order terms} beyond the time average, related to fast \textit{micromotion}, can cause heating via interactions, limiting experimental studies of many-body physics. In general, a driven system constantly exchanges energy with the driving field. Interactions redistribute this energy into other degrees of freedom, leading to an increase of the total entropy and energy. Although this heating can be suppressed in specific scenarios, e.g. via many-body localization~\cite{Zhang2017, Choi2017, Weidinger2017}, a generic closed quantum system will eventually thermalize at infinite temperature when driven~\cite{Lazarides2014}, limiting the experimental studies of many-body Floquet systems. Therefore, understanding and potentially controlling the heating in Floquet systems has triggered both theoretical~\cite{D'Alessio2014,Lazarides2014,Weidinger2017,Choudhury2014,Bilitewski2015,Choudhury2015, Genske2015} and experimental efforts~\cite{Weinberg2015,Reitter2017}.

The dynamics of a Floquet system are studied with Floquet theory. Systems heat by absorbing energy from the driving field in multiples of the energy quanta related to the modulation frequency $\omega$, caused by the scattering of the driven particles~\cite{Bilitewski2015}. The heating rate reads
\begin{equation}
\mathcal{P} = \sum_n\Gamma_n n\hbar\omega,
\label{eq.general}
\end{equation}
which is determined by the transition rates $\Gamma_n$ for the processes of absorption/emission of $n$ energy quanta. In this description, the energy exchange is quantized. 

On the other hand, in the limit of low modulation frequency where the system's intrinsic energy scales dominate, the quantization of the driving field should not have a prominent effect. The system's behavior can be described semiclassically. As a result, it is anticipated that the heating dynamics of a Floquet system have a corresponding semiclassical counterpart in this low-frequency regime. Moreover, the quantized and the semiclassical description should exhibit a continuous crossover as a function of the modulation frequency $\omega$ and amplitude.

In this work, we investigate the Floquet heating and the crossover  between the quantum and semiclassical regimes for systems subjected to periodic forcing in free space, motivated by the recent experimental demonstration of Floquet-dressed recoil momentum for photons in a two-spin mixture of cold gases~\cite{Shteynas2018} where the two spins are shaken relative to each other. Such a setting is the key ingredient of many Floquet schemes proposed for generating synthetic gauge fields and topological matter~\cite{Xu2013, Anderson2013, Yu2016,Goldman2014, Jimenez2015}. The corresponding semiclassical description of the heating in such a system is the following: the force modulates the particles' velocities and consequently generates extra kinetic energy $E_0$. This \textit{micromotion} energy $E_0$ 
can be transferred into the \textit{secular} motion of the particles via inter-particle collisions when the micromotion is out-of-phase for the colliding particles, causing an increase of the system's total energy, and consequently heating. The resulting heating rate can be estimated with the two-body elastic collision rate $\rho\sigma v$ and the associated energy $E_0$ as 
\begin{equation}
\mathcal{P}\propto \rho\sigma v E_0
\end{equation}
with the atomic density $\rho$, elastic collisional cross section $\sigma$, and the relative speed of the two particles $v$. This heating rate is continuously variable depending on the strength of the driving, characterized by $E_0$,  and is independent of the driving frequency $\omega$, which seems to contradict the Floquet description of quantized energy transfer.


In this paper, we calculate the collisional heating rates in periodically shaken atomic gases with a full Floquet treatment. We identify several distinct regimes determined by the energy hierarchies in the system and show that the semiclassical and the Floquet picture are two limiting cases of a unified general description of the heating rate as $\mathcal{P}\sim \rho\sigma v_{\rm col} E_0$. The key parameter $v_{\rm col}$ is an \textit{effective} collisional velocity parametrizing the final density of states. This can be, for example, the averaged thermal velocity $\sqrt{k_{\rm b}T/m}$ with $k_{\rm b}$ being the Boltzmann constant, $\sqrt{\hbar\omega/m}$, or $\sqrt{E_0/m}$, depending on the dominant energy scales. In addition, we show that collisional heating is suppressed in a cold fermionic system by Pauli blocking, and due to the modified density of states in systems in lower dimensions.

The paper is organized as follows: Sec.~\ref{sec.Floquet} is a concise review on the Floquet theory and scattering of Floquet-Bloch states, which serves as the theoretical basis for the main results presented in Sec.~\ref{sec.modulation}. We first analyze the Floquet heating for two atoms in Sec.~\ref{sec.two_particle_heating}. We then analyze different regimes of the collisional heating in Sec.~\ref{sec.regimes} and subsequently extend the analysis to atomic ensembles in Sec.~\ref{sec.ensemble}, including a specific discussion on fermionic systems in Sec.~\ref{sec.fermi}. A discussion of heating rates in lower-dimensional systems is presented in Sec.~\ref{sec.low}, followed by a summary and outlook in Sec.~\ref{sec.summary}.

\section{Floquet Theory and Floquet Heating}\label{sec.Floquet}

Our work is based on Floquet theory, which describes the evolution of a periodically driven system. Evolution of a Floquet system has been studied in different scenarios with different approaches, for example through high-frequency expansion~\cite{Eckardt2015,Rahav2003,Goldman2014}, Floquet-Magnus expansion~\cite{Casas2001}, and extended Hilbert space~\cite{nov2017}. We summarize here the basic concepts and formalism in Floquet theory and the scattering of the Floquet-Bloch states. This section mainly follows the description in Ref.~\cite{Bilitewski2015}; Comprehensive discussions can be found in Refs.~\cite{Bilitewski2015, Casas2001, nov2017, Goldman2014}.  

\subsection{General Aspects of Floquet Theory}
Floquet theory describes the behavior of a system governed by a time-periodic Hamiltonian $\hat{H}(t+T_0) = \hat{H}(t)$. This temporal translational symmetry allows simple descriptions of the time evolution. Solutions of the time-dependent Schr\"{o}dinger equation
\begin{equation}
\hat{H}(t)\ket{\Phi(t)} = i\hbar \partial_t \ket{\Phi(t)},
\label{eq.schrodinger}
\end{equation}
known as \textit{Floquet-Bloch states}, can be decomposed into Fourier modes as 
\begin{equation}
\ket{\Phi(t)} = \sum_l e^{-iEt/\hbar + i l \omega t} C_l \ket{\phi_l}.
\label{eq.floquet_general}
\end{equation}
Here, $\omega = 2\pi/T_0$ is the modulation frequency and $E$ is the eigenenergy of the corresponding non-driven system. The amplitude of each of the Fourier modes $C_l$ generally depends on parameters such as the strength and the frequency of the driving. 

The system does not conserve energy, due to the external drive. In the literature, two different conventions are adopted to describe the energy structure of such a system~\cite{Bilitewski2015}. Some authors define \textit{quasienergies} $E^{\rm q} = E\; {\rm mod}\ \hbar\omega$ lying between $(-\hbar\omega/2, \hbar\omega/2)$. Others distinguish between the carrier energy $E$, describing the \textit{secular motion}, and the energy sidebands $E\pm l\hbar\omega$, describing the \textit{micromotion}. This distinction can be understood by considering an adiabatic ramp of the amplitude of the driving. In this work, we adopt the second convention. 

\subsection{Scattering of Floquet-Bloch States}
The dressed energy sidebands of the Floquet-Bloch states modify the scattering between two states caused by interactions. Scattering can occur not only between the carriers but also from the carrier of the initial state to the sidebands of the final state. In the latter case, the final and the initial carrier energy are different by multiples of the energy quanta $\hbar\omega$, representing the energy exchange between the driving field and the system via scattering. This process is formulated with the so-called Floquet Fermi's golden rule~\cite{Bilitewski2015}. The transition amplitude between two Floquet-Bloch states is calculated using time-dependent purterbation theory~\cite{Bilitewski2015}:
\begin{equation}
\begin{split}
A(i\rightarrow f, t) &= -\frac{i}{\hbar}\int _0^t {\rm d}t' \bra{\Psi_{\rm f}(t')}\hat{V}\ket{\Psi_{\rm i}(t')}\\
&=-\frac{i}{\hbar}\sum_{p,q}\int _0^t {\rm d}t' e^{i[E_{\rm f}-E_{\rm i}+\hbar (p-q)\omega]t/\hbar}V^{p,q},
\end{split}
\end{equation}
where $V^{p,q}=\bra{\phi^q_{\rm f}}\hat{V}\ket{\phi^p_{\rm i}}$ is the coupling between two Fourier modes $p,q$ belonging to the final and the initial state respectively via, for example, collisions. The corresponding transition rate is readily obtained as
\begin{equation}
\begin{split}
\Gamma(i\rightarrow &f) = \lim_{t\rightarrow \infty}\frac{|A(i\rightarrow f, t)|^2}{t}\\
=&\frac{2\pi}{\hbar}\sum_{n}\left(\sum_{l,m}V^{l,l+n} V^{*m,m+n}\right)\delta(E_{\rm f}-E_{\rm i} - n\hbar\omega).\\
\end{split}
\label{eq.rate}
\end{equation}

The sum over the index $n$ explicitly reveals an important feature of the scattering between two Floquet-Bloch states. In the $n=0$ scattering channel, the initial and final states have the same carrier energy, and therefore no net energy is exchanged between the colliding particles and the driving field (\textit{Floquet elastic processes}), which resembles the conventional elastic scattering.  Scattering channels with $n\neq 0 $ characterize the processes where the energy of the atomic system is changed by exchanging $n$ quanta with the driving field (\textit{Floquet inelastic processes}), leading to Floquet heating.

\subsection{Heating Rates}
We define the heating rate $\mathcal{P}$ of the system as the rate of the average increase in the system's total energy $\dot{E}_{\rm tot}$, which can be expressed with the scattering rate $\Gamma_n$ and the related energy transfer $n\hbar\omega$ as 
\begin{equation}
\mathcal{P} = \sum_n \Gamma_n n\hbar\omega.
\end{equation}


As shown above, heating of a periodically driven system originates from the absorption of energy from the driving field through inter-particle interactions. The heating rate of a Floquet system can be calculated by first finding the exact Floquet-Bloch state wave function. Then the energy exchange rate can be obtained by calculating the transition rates for all quantized absorption/emission processes and their associated energy change. Generally, the explicit form of the wave function of the Floquet-Bloch state $\ket{\Phi(t)}$ is obtained by inserting Eq.~(\ref{eq.floquet_general}) into Eq.~(\ref{eq.schrodinger}) and solving the infinite number of coupled equations for amplitudes $C_l$. However, for some special cases, the exact solutions have a simple form, such as the system presented in Sec.~\ref{sec.modulation}.

\section{Collisional Heating for Periodic Forces}\label{sec.modulation}
We apply the method described in Sec.~\ref{sec.Floquet} to the system of interest: a spin mixture of atoms with different magnetic moments for a periodically modulated magnetic field gradient, as implemented in Ref.~\cite{Shteynas2018} (Fig.~\ref{fig.illustration}). For simplicity, we assume the two spins to have equal but opposite magnetic moments, such that they experience opposite forces. We first derive the exact analytic form of the corresponding Floquet-Bloch state wave function, then calculate the collisional heating for a single pair of atoms with opposite spins, before generalizing the results to atomic ensembles. In this section, we focus on three-dimensional systems. The results are extended to lower dimensions in Sec.~\ref{sec.low}.

\subsection{Collisional Heating for Two Atoms}\label{sec.two_particle_heating}
\subsubsection{Single-particle Floquet-Bloch states}
The Hamiltonian we consider is 
\begin{equation}
\hat{H}(t) = \frac{\hbar^2\hat{\vc{k}}^2}{2m} + \hbar k_0 \vc{z}\hat{\sigma}_{\rm z}\sin{(\omega t + \phi)},
\label{eq.hamiltonian}
\end{equation}
where the time dependent term arises from a spin-dependent periodic force $\vc{F} = \hbar k_0\sigma_{\rm z}\sin{(\omega t +\phi)}\hat{\vc{e}}_z$. 
The corresponding Floquet-Bloch states defined in Eq.~(\ref{eq.schrodinger}) have the compact and intuitive form
\begin{equation}
\begin{split}
\Psi_{\vc{k}}(\vc{r},t) = &\frac{1}{\sqrt{V}}\exp{\left[i\vc{k}(t)\cdot\vc{r} -i\Phi(t)\right]},\\
\end{split}
\label{eq.state}
\end{equation}
where
\begin{equation}
\hbar \vc{k}(t) =\hbar\vc{k}+ \hbar k_ 0\hat{\vc{e}}_z \int_0^t \sin{\omega t'}\;{\rm d}t',
\end{equation}
and
\begin{equation}
\Phi(t) = \int_0^t\frac{\hbar^2 \vc{k}(t')\cdot\vc{k}(t')}{2m}\;{\rm d}t'
\end{equation}
are the instantaneous momentum and the cumulative dynamic phase at time $t$. The physical interpretation of the wave function Eq.~(\ref{eq.state}) is made transparent by considering a stationary Gaussian wave packet $\ket{\phi}_{t = 0}= \int {\rm d}\vc{k} \exp{(-\vc{k}\cdot\vc{k}/\sigma^2_{\vc{k}})}\ket{\phi}_{\vc{k}}$ at the origin. The expectation values of the position $\vc{r}$ and the momentum $\hbar\vc{k}$ of the wave packet at time $t$ under the periodic driving
\begin{equation}
\langle \vc{r} \rangle_t =\frac{\hbar}{m}\int_0^t \vc{k}(t'){\rm d}t',\;\;\langle \hbar\vc{k}\rangle_t=  \hbar \vc{k}(t)
\end{equation}
are identical to those of a driven classical particle. We further identify the \textit{secular} motion of the particle with the time average of $\langle \vc{r} \rangle_t,\langle \hbar\vc{k}\rangle_t$ over a period $T_0$.

The periodic modulation at the driving frequency $\omega$ appears in both the dynamic phase $\Phi(t)$ and the wave vector $\vc{k}(t)$. The amplitude $C_l$ of each Fourier mode defined in Eq.~(\ref{eq.floquet_general}) can be readily obtained via the expansion $e^{ia\sin \omega t} = \sum_{n= -\infty}^{\infty}J_n(a)e^{i n \omega t}$ as $C_l = \sum_{i+j+2k = l} J_i(k_0 z) J_j(\alpha) J_k(\beta)$ with the First-order Bessel functions $J_{\nu}$ and two parameters defined as
\begin{equation}
\alpha_{\vc{k}} = \frac{\hbar k_z k_0}{m\omega},\;\beta = \frac{\hbar  k^2_0}{8m\omega}.
\end{equation}
These motional sidebands dressed by the periodic driving have been directly observed via resonant fluorescence spectroscopy in trapped-ion systems~\cite{Raab2000}. The result is also conceptually similar to an optical modulator where the carrier frequency is dressed with frequency sidebands due to the periodic modulation of the medium's optical properties.
\begin{figure}[h]
\includegraphics[width = 8.34cm]{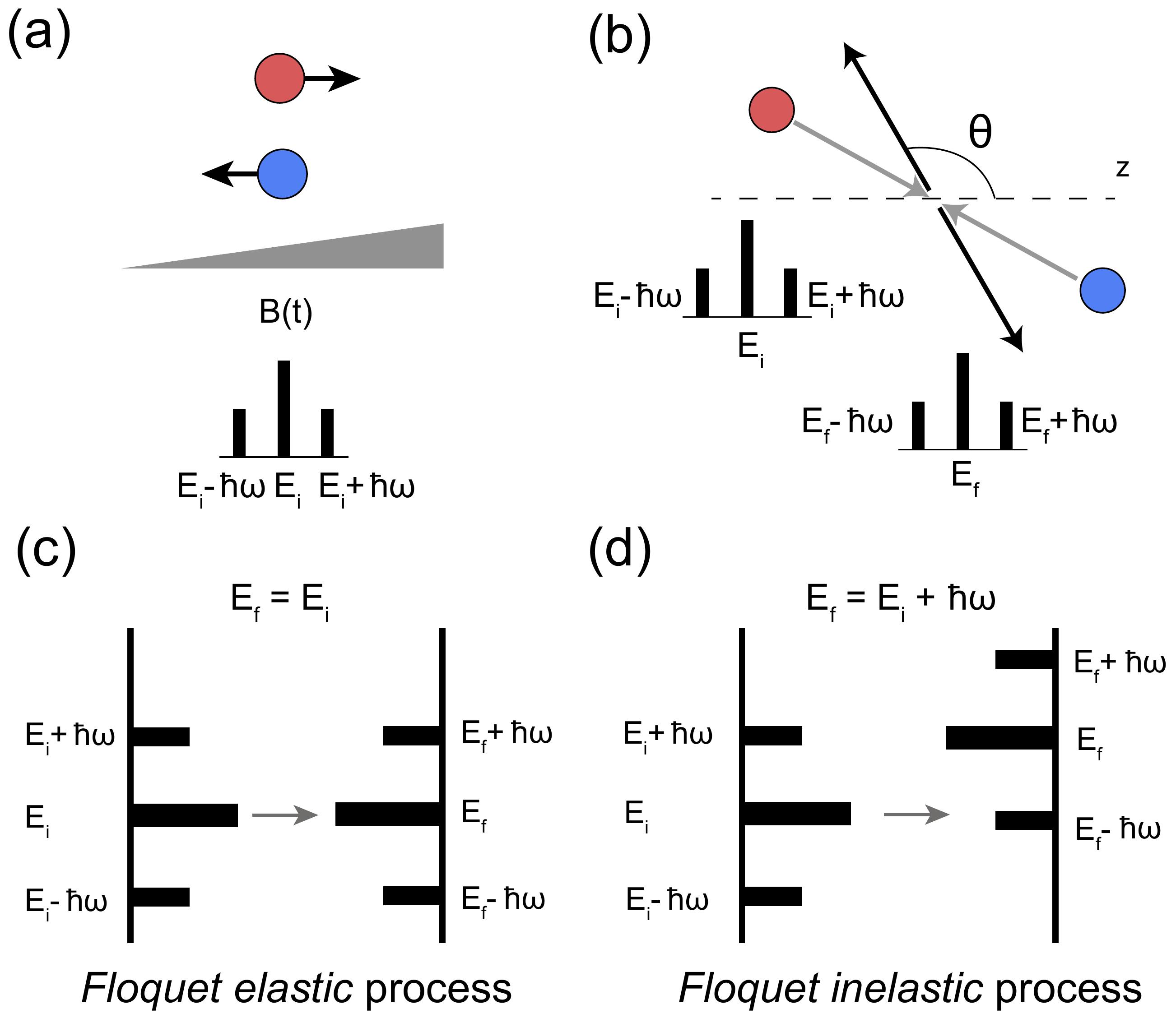}
\caption{\label{fig.illustration} Scattering between two particles subjected to a periodic force. Atoms with different spins are represented by red and blue. a) The oscillating force dresses each particle with energy sidebands spaced by $\hbar\omega$. The scheme has been implemented in Ref.~\cite{Shteynas2018} where atoms with different magnetic moments are driven by a periodic magnetic field gradient. b) Elastic collisions couple two Floquet-Bloch states. Gray and black arrows represent the incoming and the outgoing states respectively and $\theta$ is the scattering angle. c - d) Illustration of the \textit{Floquet elastic} (c) and \textit{Floquet inelastic} (d) process. Besides the regular elastic collisions where the final and the initial state have the same carrier energy (\textit{Floquet elastic process}), the existence of the energy sidebands allows transitions between states whose carrier energies are different by a multiple of $\hbar \omega$ (\textit{Floquet inelastic process}), leading to the exchange of energy between the system and the driving field. }
\end{figure}
\subsubsection{Two-particle collisions and the heating rates}\label{sec.collisions}
The two-body problem is reduced to a single particle problem by decomposing the dynamics into relative and center-of-mass parts. The center-of-mass motion is unaffected by collisions and is therefore omitted in further calculations. The two-body Hamiltonian reads
\begin{equation}
\begin{split}
\hat{H}(\vc{R},\vc{r},t) =&\frac{\hbar^2\hat{\vc{K}}^2}{2M} +\frac{\hbar^2\hat{\vc{k}}^2}{2\mu} + \hbar k_0 \hat{\sigma}_z\vc{z} \sin{(\omega t + \phi)}\\
&+g\delta(\vc{r}),
\end{split}
\label{eq.hamiltonian_relative}
\end{equation}
with $\mu = m/2$ being the reduced mass. $\vc{r} =( \vc{r}_1-\vc{r}_2)/2$, $\vc{k} =( \vc{k}_1-\vc{k}_2)/2$ are the relative coordinate and momentum. 

The wave functions for the relative motion have the same structure as Eq.~(\ref{eq.state}), except that the mass is replaced by the reduced mass $\mu$ and the momentum by the relative momentum $\hbar\vc{k}$.

Collisions are captured by the $s$-wave pseudopotential $\hat{V} = g\delta(\vc{r})$  described by Eq.~(\ref{eq.hamiltonian_relative}). Here $g = 4\pi\hbar^2a/m$ is the strength of the interaction and $a$ is the $s$-wave scattering length between the two spin states. The interaction $\hat{V}$ couples two Floquet-Bloch states in Eq.~(\ref{eq.state}). Without the periodic driving, elastic collisions couple only states with the same kinetic energy $ E_{\vc{k}} = E_{\vc{k'}}=\hbar^2 |\vc{k}|^2/2\mu $. However, the energy sidebands introduced by the periodic driving, formulated in Eq.~(\ref{eq.state}), allow the scattering between states whose carrier energies differ by a multiple of $\hbar\omega$, giving $E_{\vc{k}'} - E_{\vc{k}}=n\hbar\omega$ (Fig.~\ref{fig.illustration}(b)-(d)). The associated quantized energy change $n\hbar\omega$ is transferred to the secular motion, leading to heating (or cooling). The transition rate from the ingoing state $\ket{\Psi_{\vc{k}}^{\rm i}}$ to the outgoing state $\ket{\Psi_{\vc{k'}}^{\rm f}}$ can be readily calculated from Eq.~(\ref{eq.rate}). By combining Eqs.~(\ref{eq.rate}) and (\ref{eq.state}), we derive the coupling matrix element
\begin{equation}
\begin{split}
M_n(\vc{k}\rightarrow \vc{k'}) =&g\sum_l \int {\rm d}\vc{r}\;\delta(\vc{r})C_l(\vc{k})^* C_{l+n}(\vc{k'})\\=&g J_n(\alpha_k - \alpha_{k'}),\\
\end{split}
\end{equation}
which gives the total transition rate from $\ket{\Psi_{\vc{k}}^{\rm i}}$ to  $\ket{\Psi_{\vc{k'}}^{\rm f}}$
\begin{equation}
\begin{split}
&\Gamma(\vc{k}\rightarrow \vc{k}') \\&=\sum_n\Gamma_n(\vc{k}\rightarrow \vc{k}')\\&=\sum_n \frac{2\pi}{\hbar} |M_n(\vc{k}\rightarrow \vc{k'})|^2 \delta (E_{\vc{k'}}-E_{\vc{k}}-n\hbar\omega),
\end{split}
\end{equation}
explicitly showing the scattering rate of channels with different numbers of energy quanta $\hbar\omega$ exchanged. The scattering matrix element $M_n$ reveals the microscopic process of the energy exchange with the driving field ($n\neq 0 $ processes): it occurs only when  $\alpha_k \neq \alpha_{k'}$, i.e. when the projection of the relative momentum $\vc{k}$ to the shaking axis changes.

\begin{figure}[h]
\includegraphics[width = 8.5cm]{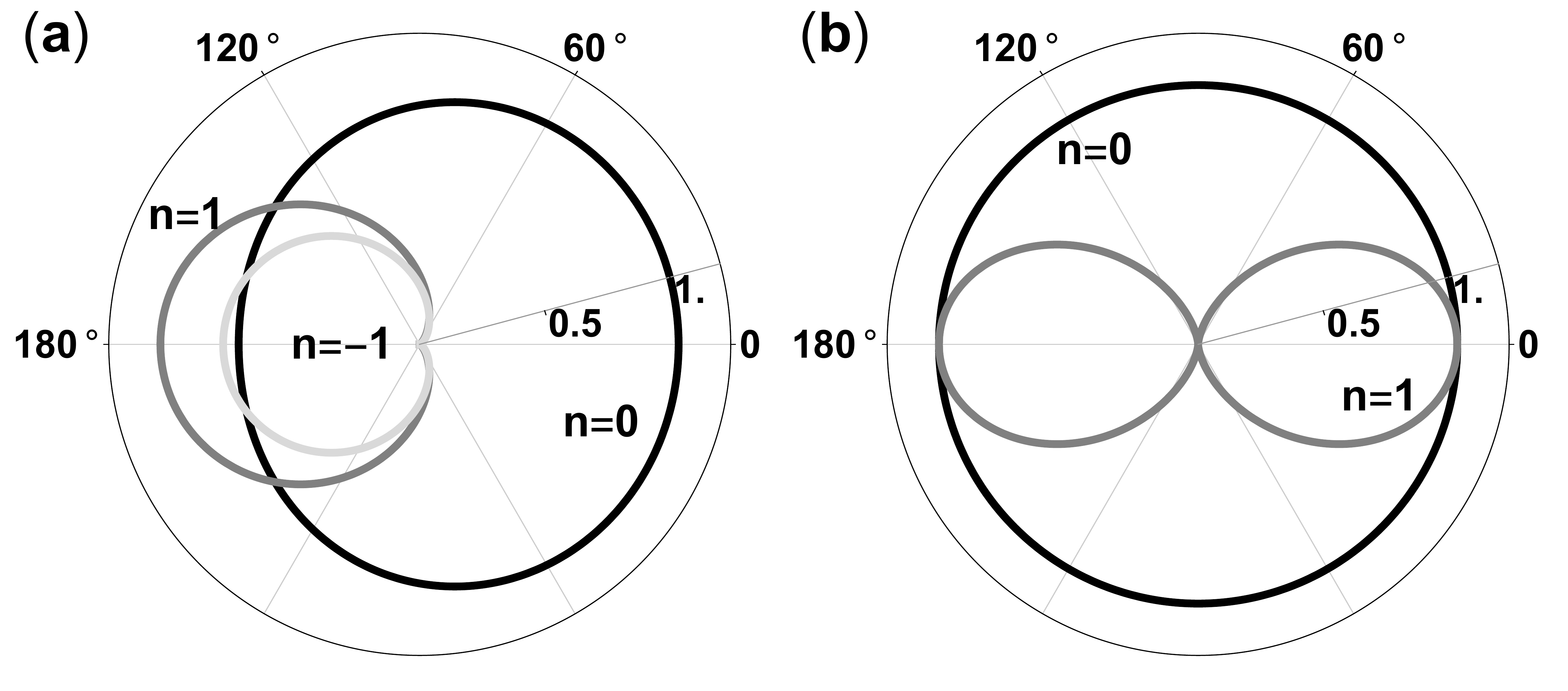}
\caption{\label{fig.cross_sec} Differential scattering cross sections for Floquet scattering processes with $n=0$(black), $n = 1$(gray), and $n=-1$(light gray). The angular coordinate corresponds to $\theta$ as depicted in Fig.~\ref{fig.illustration}. (a) For $E_{\vc{k}} > \hbar\omega$, both absorption ($n >0$) and emission ($n < 0$) processes are allowed. Scattering between the Floquet-Bloch states is anisotropic in angle due to the sidebands. The figure is plotted for $\alpha = 0.4, \beta = 0.003$. (b) For strong driving the micromotion dominates, and the forward and backward scattering are symmetric as expected. The maximum scattering cross sections are normalized to unity, except for the process with $n=-1$ in (a) which is normalized with the maximal cross section of the $n = 1$ process in (a).}
\end{figure}
A feature of the scattering between two dressed particles is the anisotropy in the scattering cross section, as shown in the differential scattering cross section (Fig.~\ref{fig.cross_sec})
\begin{equation}
\frac{{\rm d}^2\sigma_n}{{\rm d}\Omega\;{\rm d}E} = \frac{2\sigma}{\sqrt{m}}\frac{|\vc{k}'|}{|\vc{k}|}|M_n|^2\delta (E_{\vc{k'}}-E_{\vc{k}}-n\hbar\omega).
\end{equation}
Though the potential $\hat{V}$ is isotropic, the scattering cross sections are \textit{anisotropic} for each channel. This anisotropy in the scattering could be potentially observed in the time-of-flight pattern of a spin-mixed driven condensate. 

We calculated the heating rate $\mathcal{P} = \sum_n \Gamma_n n\hbar\omega$ for a single pair of colliding particles with a relative momentum $\hbar\vc{k}$ by summing over the allowed final states $\vc{k}'$ and scattering channels $n$, leading to 
\begin{equation}
\begin{split}
\mathcal{P}_{\vc{k}} =& \sum_n \sum_{\vc{k}'}  \Gamma_n (\vc{k}\rightarrow \vc{k'}) n \hbar \omega\\
=&\sum_n\frac{2\pi }{\hbar}\frac{g^2}{V}D_{\rm 3D}(E_{\vc{k}}+n\hbar\omega)\gamma^2(\vc{k},n)n\hbar\omega, 
\end{split}
\label{eq.general_heating}
\end{equation}
where
\begin{equation}
\begin{split}
&\gamma^2(\vc{k},n) = \\
&\frac{1}{2g^2}\int_0^{\pi} {\rm d}\theta\; \sin\theta|M_n(|\vc{k}|\rightarrow\sqrt{|\vc{k}|^2+2\mu n\omega/\hbar}\cos\theta)|^2
\end{split}
\end{equation}
characterizes the transition amplitude, and $D_{\rm 3D}(E) = (2\mu/\hbar^2)^{3/2}\sqrt{E}/(2\pi)^2$ is the three-dimensional density of states of a free particle with energy $E$. Here we have assumed for simplicity that the initial relative momentum $\vc{k}$ is along the direction of modulation.
\subsection{Regimes and Crossovers}\label{sec.regimes}
One of the major results of this paper is to show the connection between the Floquet picture, where energy transfer is quantized, and the semiclassical picture, where the energy transfer is continuous. The consolidation of the two pictures can be demonstrated already by examining the two-particle calculation presented above. 


We recognize three fundamental energy scales in the system: 1) $E_0 = \hbar^2k_0^2/(4m)$ characterizing the micromotion, and the strength of the modulation. 2) $\hbar\omega$ characterizing the modulation quanta, and 3) $E_{\vc{k}}$ characterizing the relative motion between the two colliding particles, e.g. $k_{\rm b}T$ for a thermal system or $E_{\rm F}$ for a cold Fermi gas. The heating behavior of the system is qualitatively different depending on the relationship between these quantities. 

We identify the following regimes from Eq.~(\ref{eq.general_heating}) : 
\paragraph{Rapid-modulation Regime}$\hbar\omega \gg E_{\vc{k}},E_0$: A system with this condition has three features. First, only energy absorption is allowed. Second, the energy of the final states, $E_{\vc{k}}+\hbar\omega$, is now dominated by the modulation energy and $D_{\rm 3D}(E_{\vc{k}}+\hbar\omega)\sim \sqrt{\hbar\omega}$. Finally, since $M_n \sim 1/\omega^{n/2}$, the transition rate for the multi-quanta processes scales with $1/\omega^{n}$ and is therefore negligible. The heating rate can be obtained analytically by considering only the $n = + 1$ process and reads
\begin{equation}
\mathcal{P}_\vc{k} \approx \frac{4}{3} \frac{\sigma}{V}\sqrt{\frac{2\hbar\omega}{\mu}} E_0.
\label{eq.heating_bec}
\end{equation}
This is the regime where the quantum and the semiclassical picture diverge. Though the amplitudes of the sidebands drop with increasing modulation frequency, the system's heating rate increases due to the larger final density of states and the energy transfer $\hbar\omega$. 

\paragraph{Semiclassical Regime} $E_{\vc{k}}\gg \hbar\omega, E_0$: In this regime, as realized in Ref.~\cite{Shteynas2018}, the final density of states is approximated to be $D_{\rm 3D}(E_{\vc{k}}+n\hbar\omega) \approx \sqrt{E_{\vc{k}}}[1+n\hbar\omega/(2E_{\vc{k}})]$. Both the energy absorption (heating) and emission (cooling) processes are allowed. The heating of the system comes from the imbalance between absorption and emission of energy quanta $\hbar\omega$, due to the higher density of states and the larger value of the scattering matrix element for the energy absorption process. If a stronger criterion $E_{\vc{k}}\gg \hbar \omega\gg \sqrt{E_0E_{\vc{k}}}$ is fulfilled such that $\alpha_{k} \ll 1$, only sidebands with $l = \pm 1$ are relevant. In this case, we obtain from Eq.~(\ref{eq.general_heating})
\begin{equation}
\mathcal{P}_{\vc{k}} \approx 8\frac{\sigma}{V}\sqrt{\frac{2E_{\rm \vc{k}}}{\mu}}E_0,
\label{eq.heating}
\end{equation}
showing the same dependence on parameters as the semiclassical picture. The heating of the system can be understood in the semiclassical picture where the collision rate is proportional to the initial relative velocity and the modulation energy gets transferred as heat to the secular motion.

Contributions from multi-quanta transfer processes $|n|>1$ can be important. Indeed, as shown in Fig.~\ref{fig.3dheating}, results obtained with the single-quantum transfer assumption deviate at small $\omega$ from the results where higher energy transfer processes are considered. However, the heating rate at lower $\omega$ with all the higher energy transfer processes included still converges to the semiclassical limit Eq.~(\ref{eq.heating}) obtained from the single-sideband approximation. 

\paragraph{Strong-drive Regime} $E_0 \gg \hbar\omega, E_{\vc{k}}$: In this limit, the strongly driven oscillation dominates over the particle's initial motion. The scattered particles, therefore, behave as if each particle were moving at velocity $\sqrt{E_0/m}$.  Multi-quanta processes contribute significantly to the heating rate due to the large modulation index $\beta\sim \sqrt{E_0/(\hbar\omega)}$ of the final state. The heating rate reads
\begin{equation}
\begin{split}
\mathcal{P}_{\vc{k} = 0}=&\frac{2\pi}{\hbar}g^2\sum_{n=0}^{\infty}\bigg[D_{\rm 3D}(n\hbar\omega)n\hbar\omega \\
&\times\int d\theta \frac{\sin{\theta}}{2}\;|J_n\left(4\sqrt{\frac{nE_0}{\hbar\omega}}\cos{\theta}\right)|^2\bigg]\\
=&3.36\;\sigma\sqrt{\frac{E_0}{m}}E_0,
\end{split}
\label{eq.strong}
\end{equation}
where the coefficient $3.36$ is found numerically.

\subsection{Ensemble Heating Rates}\label{sec.ensemble}
We now apply the two-particle results above to thermal ensembles with total atom number $N$ by averaging over the ensemble as
\begin{equation}
\begin{split}
\mathcal{P}_{\rm ens} =& \int \frac{{\rm d}^3\vc{k}_1}{(2\pi)^3}\frac{{\rm d}^3\vc{k}_2}{(2\pi)^3}\frac{{\rm d}^3\vc{k}'_1}{(2\pi)^3}\frac{{\rm d}^3\vc{k'}_2}{(2\pi)^3}\delta_{\vc{k}_1+\vc{k}_2,\vc{k}'_1+\vc{k}'_2}\\
&\times f(\vc{k}_1)f(\vc{k}_2)\mathcal{P}(\vc{k}_1,\vc{k}_2\rightarrow\vc{k}'_1,\vc{k}'_2),
\end{split}
\label{eq.ensemble}
\end{equation}
where $f({\vc{k}})$ is the particles' velocity distribution and $\mathcal{P}(\vc{k}_1,\vc{k}_2\rightarrow\vc{k}'_1,\vc{k}'_2) = \mathcal{P}(\vc{k}\rightarrow\vc{k'})$ with $\vc{k} = (\vc{k}_1-\vc{k}_2),\vc{k'}=(\vc{k}'_1-\vc{k}'_2)$. We calculate the heating rates at different regimes for 1) thermal clouds at temperature $T$, 2) a Bose-Einstein Condensate, and 3) a degenerate Fermi gas.

\begin{figure}[h]
\includegraphics[width = 8.34cm]{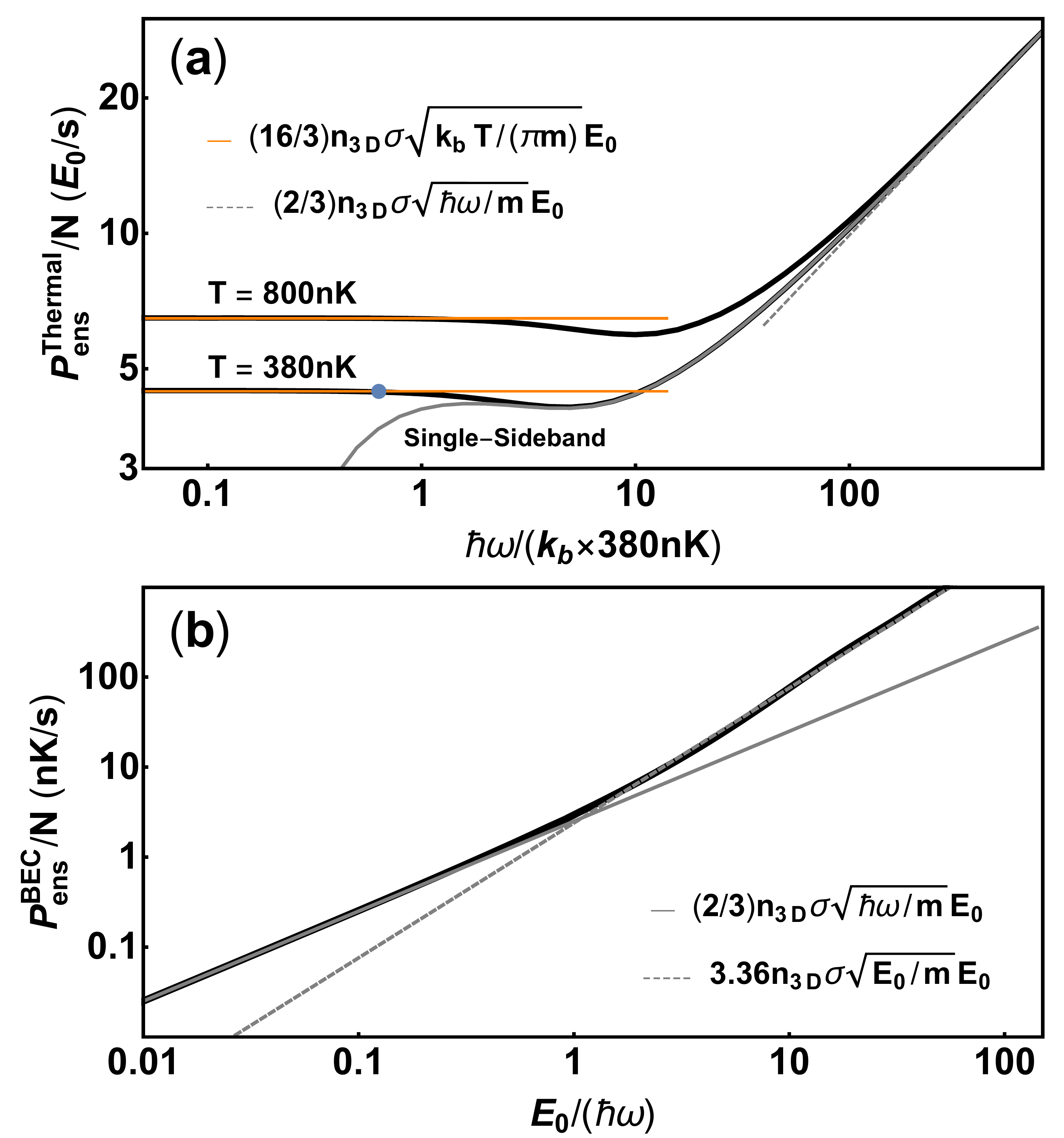}
\caption{\label{fig.3dheating} Numerical calculations of the heating rate involving multi-quanta process $|n|\geq 1$ for a thermal cloud at temperature $T=380\;{\rm nK}$ (experimental condition in~\cite{Shteynas2018}), $800\;\rm nK$ and for $T=0$ (i.e. an ideal Bose-Einstein condensate) at $\omega = 2\pi\times 5\;\rm kHz$. a) For the thermal cloud, the heating rates scale with $\sqrt{\omega}$ at high frequencies and become independent of $\omega$ around $\hbar \omega \approx 4k_{\rm b} T$ which is the ensemble averaged $E_{\vc{k}}$. The heating rates, then plateau at the semiclassical value given by Eq.~(\ref{eq.heating_thermal}), illustrated by the orange lines. The solid gray and the dashed gray lines represent the results obtained from the single-sideband approximation Eq.~(\ref{eq.heating_fast}) for $T = 380\;\rm nK$.  
The blue dot indicates the parameter implemented in~\cite{Shteynas2018}. b) For a condensate, the heating rate scales quadratically with the modulation strength characterized by $E_0$ when $\hbar\omega \gg E_0$ and shows semiclassical behavior at $\hbar\omega \ll E_0$. The black line is the numerical result, the dashed gray line is the heating rate calculated with the single-sideband approximation which matches the numerical result in the regime $\hbar\omega \gg E_0$. In this calculation we use $n_{\rm 3D} = 10^{12}\;{\rm cm}^{-3}$ and scattering length $a =53.8 a_0$.}
\end{figure}

The analytic results presented in this section are obtained by assuming the single-sideband approximation unless otherwise stated, as justified in the previous Section. Multi-quanta results are presented as numerical results in Fig.~\ref{fig.3dheating}. 

\paragraph{Thermal Ensemble} For a classical ensemble at a temperature $T$, the distribution $f(\vc{k}) $ is the Boltzmann distribution. The resultant ensemble heating rate is
\begin{equation}
\mathcal{P}^{\rm Thermal}_{\rm ens}=\frac{2}{3} Nn_{\rm 3D}\sigma\sqrt{\frac{\hbar\omega}{m}} E_0 
\label{eq.heating_fast}
\end{equation}
for the \textit{rapid-modulation regime}, and 
\begin{equation}
\mathcal{P}^{\rm Thermal}_{\rm ens}=\frac{16}{3} Nn_{\rm 3D}\sigma\sqrt{\frac{kT}{\pi m}}E_0\\
\label{eq.heating_thermal}
\end{equation}
for the \textit{semiclassical regime}. Along with the general expression $\mathcal{P}\sim \rho\sigma v_{\rm col} E_0$, we identify $v_{\rm col} \propto \sqrt{k_{\rm b}T/m}$ to be the ensemble averaged thermal velocity, which reproduces the semiclassical picture where the micromotion energy is transferred to the secular motion via elastic collisions at a rate proportional to the thermally averaged relative speed between the two colliding particles.

\paragraph{Bose-Einstein Condensates}A Bose-Einstein condensate at $T = 0 $ is treated as an ensemble with $f(\vc{k}) = \delta(0)$. Following Eqs.~(\ref{eq.heating_bec}) and~(\ref{eq.strong}), the heating rate reads
\begin{equation}
\mathcal{P}_{\rm ens}^{\rm BEC}=3.36Nn_{\rm 3D}\sigma\sqrt{\frac{E_0}{m}}E_0
\end{equation}
at $E_0 \gg \hbar\omega$ and 
\begin{equation}
\mathcal{P}_{\rm ens}^{\rm BEC}=\frac{2}{3}Nn_{\rm 3D}\sigma\sqrt{\frac{\hbar \omega}{m}}E_0
\end{equation}
at $E_0 \ll \hbar\omega$, as shown in Fig.~\ref{fig.3dheating}(b). The result suggests a collisional velocity $v_{\rm col} \propto \sqrt{E_0/m}$ and $v_{\rm col}\propto \sqrt{\hbar\omega/m}$ respectively.

To compare with the experimental results in Ref.~\cite{Shteynas2018}, we numerically calculate the heating rates for various cases by directly calculating the sum Eq.~(\ref{eq.general_heating}) for the experimental parameters (Fig.~\ref{fig.3dheating}). Our result is consistent with the weak Floquet heating observed. 
\subsection{Fermionic Systems}\label{sec.fermi}
The heating effect studied in this work relies on atomic collisions which can be affected by particle statistics. For deeply degenerated Fermi gases, Pauli blocking effectively reduces the elastic collisional cross section $\sigma$, as experimentally demonstrated in~\cite{DeMarco2001,Ferrari1999,Holland2000}. As a result, collisional heating from periodic driving is suppressed in a fermionic system.

Fermi statistics dominates when $E_{\rm F}$ is the largest energy scale. At this condition, collisions occur on the Fermi surface. More formally, the heating rate for a fermionic system is expressed as
\begin{equation}
\begin{split}
\mathcal{P}_{\rm ens}^{\rm F}= &\int \frac{{\rm d^3}\vc{k}_1}{(2\pi)^3} \frac{{\rm d^3}\vc{k}_2}{(2\pi)^3}\frac{{\rm d^3}\vc{k_1'}}{(2\pi)^3} \frac{{\rm d^3}\vc{k_2'}}{(2\pi)^3}\delta_{\vc{k}_1+\vc{k}_2,\vc{k}'_1+\vc{k}'_2}\\
&\times f_{\downarrow}(\vc{k}_1)f_{\uparrow}(\vc{k}_2)[1-f_{\downarrow}(\vc{k}'_1)][1-f_{\uparrow}(\vc{k}'_2)]\\
&\times\mathcal{P}(\vc{k}_1,\vc{k}_2\rightarrow\vc{k}'_1,\vc{k}'_2).\\
\end{split}
\end{equation}
Here, $f =1/(g_i + 1)$ is the occupation number with $g = \exp{[(\hbar^2\vc{k}\cdot\vc{k}/(2m)-\mu)/k_{b}T]}$, and $\mu$ is the chemical potential. Pauli blocking is captured by the extra factor $(1-f_{\downarrow}(\vc{k}'_1))(1-f_{\uparrow}(\vc{k}'_2))$, accounting for the occupation of the final states. Here we consider an equal mixture of spin-up and spin-down atoms. 

When $\hbar\omega$ is the largest energy scale in the system, the entire Fermi sea is involved in the collisional heating process since all possible final states are unoccupied. The heating rate scales with $\omega$ in a similar way as in the case where Fermi statistics is absent as in Eq.~(\ref{eq.heating_fast}).

When $\hbar\omega < E_{\rm F}$, Pauli blocking occurs. At $T = 0$, $f$ approaches a step function with $\mu = E_{\rm F}$. Collisions occur in a shell with a thickness of $\sim (\hbar\omega/E_{\rm F})k_{\rm F}$ around the sharp Fermi surface. We show in Appendix \ref{app.fermi} that the heating rate of the 3D system becomes
\begin{equation}
\mathcal{P}_{\rm ens}^{F}\approx \frac{\pi}{ \sqrt{2}}Nn_{\rm 3D}\sigma \left(\frac{\hbar\omega}{E_{\rm F}}\right)^2\sqrt{\frac{E_{\rm F}}{m}}E_0
\label{eq.fermi_3D}
\end{equation} 
when $\hbar\omega < E_{\rm F}$, where the factor $(\hbar\omega/E_{\rm F})^2$ characterizes the effect of Pauli blocking. The power law of $\hbar\omega$ originates from three effects: 1) scatterings occurs in a shell at the surface of the Fermi sphere, accounting for a factor of $(\hbar\omega/E_{\rm F})^3$, 2) the scattering matrix elements contribute $1/(\hbar\omega)^2$, 3) the energy transfer per Floquet inelastic scattering process gives $\hbar\omega$.

\begin{figure}[h]
\includegraphics[width = 8.34cm]{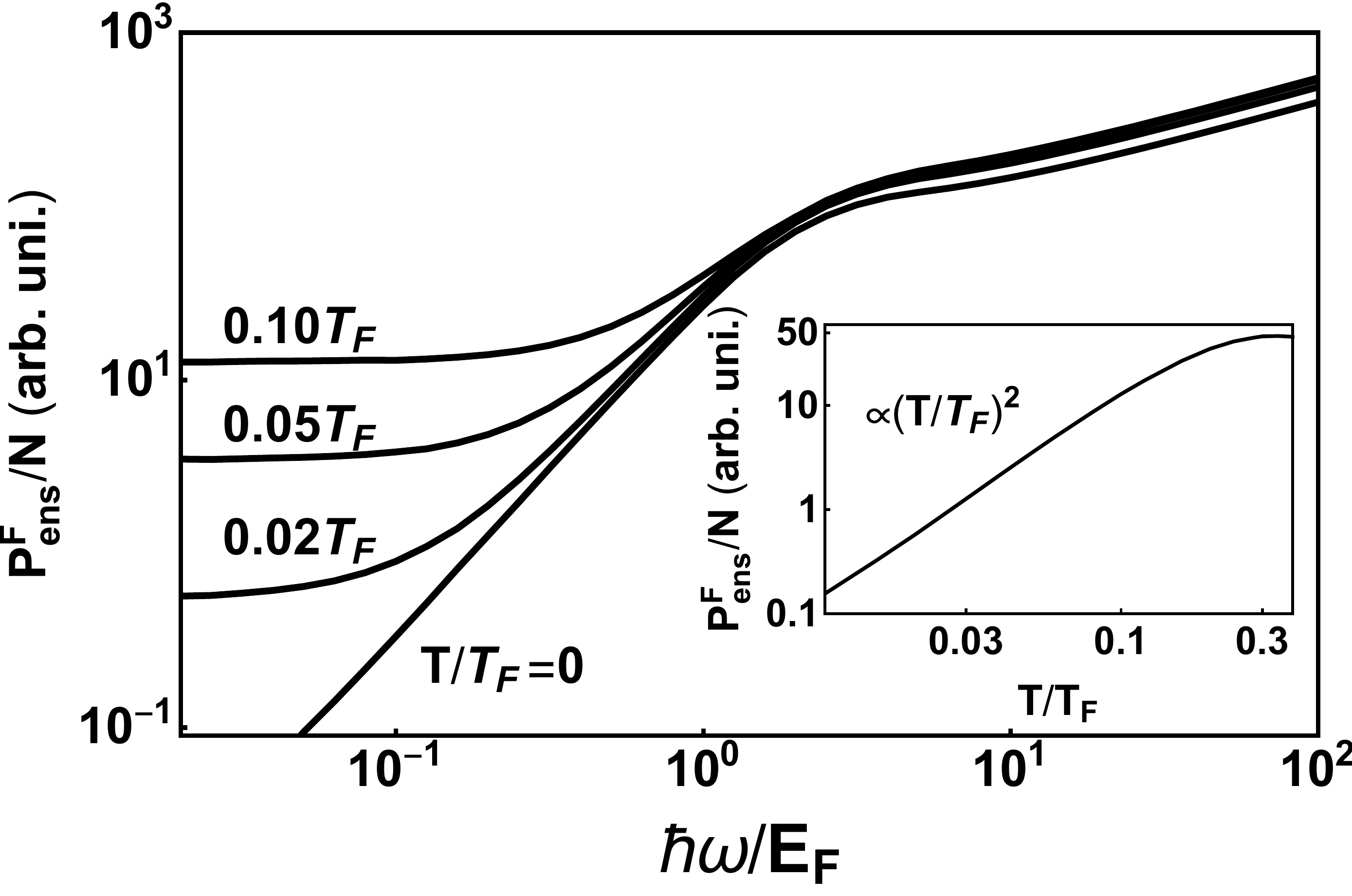}
\caption{\label{fig.fermi} Collisional heating rates in a periodically driven spin mixture of degenerated Fermi gases at $T = 0.1T_{\rm F},\;0.05T_{\rm F},\;0.02T_{\rm F}$. The heating effects are Pauli suppressed for a cold Fermi gas when the energy quantum $\hbar\omega$ is smaller than the Fermi energy $E_{\rm F}$. The discrepancies at high modulation frequencies are numerical artifact. Inset: heating rates at $\hbar\omega = 0.016\;E_{\rm F}$  and fixed $E_{\rm F}$ for various temperatures where $\hbar\omega < k_{\rm b}T < E_{\rm F}$. The heating rate is proportional to $(T/T_{\rm F})^2$.}
\end{figure}

For $T\neq 0$, thermal excitations smear the Fermi surface, affecting the number of states involved in the collisional processes. When $k_{\rm b}T <\hbar\omega$, the modulation energy still dominate the scattering. The result is similar to the $T = 0$ cases, as shown in Fig.~\ref{fig.fermi}.
When $k_{\rm b}T > \hbar\omega$,  the thickness of the collisional shell in momentum space is on the order of $(k_{\rm b}T/E_{\rm F})k_{\rm F}$. We show in Appendix \ref{app.fermi} and also numerically that the heating rate scales as 
\begin{equation}
 \mathcal{P}_{\rm ens}^{F}\propto Nn_{\rm 3D}\sigma\left( \frac{T}{T_{\rm F}}\right)^2\sqrt{\frac{E_{\rm F}}{m}} E_0
\end{equation}
at low temperature and is independent of the modulation frequency.

When the micromotion energy $E_0$ is large compared with both the thermal energy $k_{\rm b}T$ and the modulation energy $\hbar\omega$, the heating rate of the system becomes
\begin{equation}
 \mathcal{P}_{\rm ens}^{F}\propto Nn_{\rm 3D}\sigma\left( \frac{E_0}{E_{\rm F}}\right) \sqrt{\frac{E_{\rm F}}{m}} E_0,
\end{equation}
which can be explained by considering the collision between two Fermi spheres displaced by $k_0$ from each other. Collisions are only allowed within a Fermi shell with a thickness of $k_0$. The number of available states is therefore $\propto \left(k_0/k_{\rm F}\right)^2N^2$ which implies the Pauli blocking factor $\left(E_0/E_{\rm F}\right)^2$.

\subsection{Lower-Dimension Systems}\label{sec.low}
Though our calculations are done for a specific system,  several conclusions are generally valid. The result that $\mathcal{P}\sim \sqrt{\hbar\omega}$ at high frequencies is valid for any three-dimensional systems in free space with quadratic particle dispersion due to the density of states. Since the Floquet \textit{elastic} collisional rate is bounded, such Floquet systems \textit{cannot} be studied in thermal equilibrium in the limit of fast modulation frequencies. However, it is often desirable to have the modulation energy scale $\hbar\omega$ greater than all the other dynamic energy scales. 

One possible solution is to go to lower dimensions, suggested by the observation that heating at high frequency originates from the increased density of states with higher energy $D_{\rm 3D}\sim \sqrt{\hbar\omega}$.  In a 2D system, the density of states is independent of the energy, so that excessive collisional heating can be suppressed. 

We consider here quasi-2D scenarios where the atomic motion is constrained in a two-dimensional pancake but scattering is still described by the 3D $s$-wave pseudopotential. This can be achieved in the case where the scattering length $a$ is much larger than the interaction range but smaller than the oscillator length  $l_0 = \sqrt{\hbar/(m\nu_{\perp})}$ in the strongly confined direction, with trapping frequency $\nu_{\perp}$. The modulation is in-plane.

In these cases, the Floquet-Bloch wave function is written as
\begin{equation}
\Psi(x, {\pmb{\rho}},t) = \phi_{\perp}(x)\frac{1}{\sqrt{A}}\exp{\left[i\vc{k}(t)\cdot\pmb{\rho} -i\Phi(t)\right]},
\label{eq.2d_wf}
\end{equation}
where $\pmb{\rho} = \{y , z\}$ and $\vc{k} = \{k_y,k_z\}$ are the 2D radial vector and wave vectors. $A$ is the system area. The component along the strongly confined direction $x$ has been explicitly separated as $\phi_{\perp}$. We assume that $\phi_{\perp}(x) = \pi^{1/4}l_0^{1/2}\exp{(-x^2/2l_0^2)}$ and particles stay in the ground state wave function.

With these parameters, results obtained for the 3D case can be readily extended to quasi-2D by replacing the scattering strength $g$ with the effective 2D scattering strength~\cite{Petrov2000}
\begin{equation}
g_{\rm 2D} =  \frac{2\sqrt{2\pi}\hbar^2}{m}\frac{a}{l_0}.
\end{equation}
Together with the 2D density of states $D_{\rm 2D} = \mu/(2\pi\hbar^2 )$ and $\sigma = 4\pi a^2$, the heating rate can be expressed as (see Appendix~\ref{app.2d} for details)
\begin{equation}
\mathcal{P}_{\rm 2D} =16 Nn_{\rm 2D}\frac{\hbar}{m}\frac{\sigma}{l_0^2}\left(\frac{8k_{\rm b}T}{\hbar\omega}+ 1\right)E_0
\label{eq.2dfast}
\end{equation}
in the \textit{rapid-modulation regime} and
\begin{equation}
\mathcal{P}_{\rm 2D}= 32 N n_{\rm 2D}\frac{\hbar}{m}\frac{\sigma}{l_0^2}E_0
\label{eq.2dsemi}
\end{equation}
in the \textit{semiclassical regime}. This can be interpreted as a 3D density $n_{\rm 3D} = n_{\rm 2D}/l_0 = N/(Al_0)$, and a velocity $v_{\rm col} = \hbar/(m l_0)$. The heating rates are bounded in both regimes.
\begin{figure}[h]
\includegraphics[width = 8.34cm]{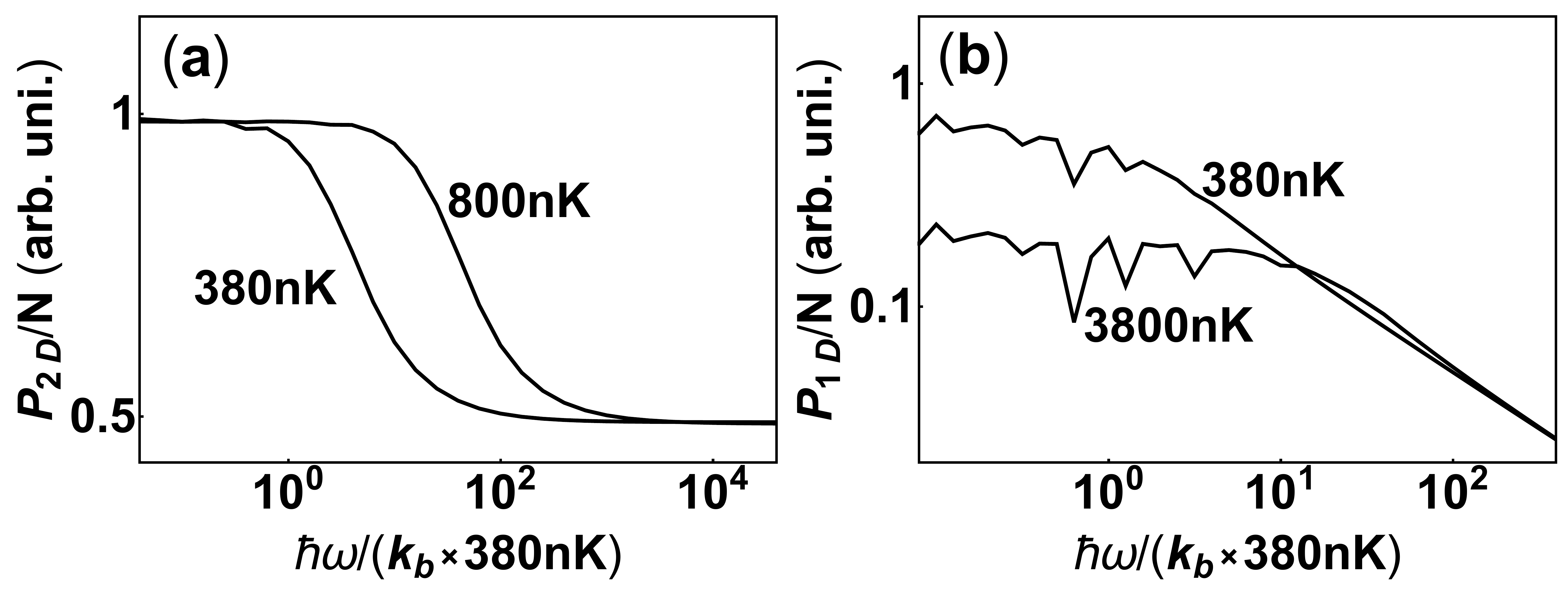}
\caption{\label{fig.2D} Numerical calculations for the collisional heating rates in a quasi-2D (a) and quasi-1D system (b). The heating rate is independent of $\omega$ at high modulation frequencies in quasi-2D regime, and is suppressed with the increasing modulation frequency $\hbar\omega$ in quasi-1D. The oscillations in (b) are numerical artifacts.}
\end{figure}

Similarly, for a quasi-1D system with length $L$,  $g_{\rm 1D} = \hbar^2a/(ml^2_0)$ is the interaction strength. We obtained
\begin{equation}
\begin{split}
&\mathcal{P}_{\rm 1D} \propto N n_{\rm 1D}\left(\frac{\hbar}{m }\right)^2\sqrt{\frac{m}{\hbar \omega}}\frac{\sigma}{l_0^4} E_0\;\;\; (\hbar\omega\gg k_{\rm b}T),\\
&\mathcal{P}_{\rm 1D} \propto N n_{\rm 1D}\left(\frac{\hbar}{m }\right)^2\sqrt{\frac{m}{k_{\rm b}T}}\frac{\sigma}{l_0^4}  E_0\;\;\; (\hbar\omega\ll k_{\rm b}T),\\
\end{split}
\label{eq.1d}
\end{equation}
with $n_{\rm 1D} = N/L$. The heating is now suppressed at high frequencies. 

However, we note that the modulation frequency $\omega$ is assumed to be smaller than the trapping frequency along the direction of the strong confinement $\omega \ll \nu_{\perp}$ to avoid excitations to higher oscillation states. In the opposite regime, systems are expected to reduce to the 3D case. These effects are addressed in the studies of collisional heating in modulated optical lattices~\cite{Choudhury2014, Reitter2017,Choudhury2015}, which is beyond the scope of this work. 

\section{Discussions and Summary}\label{sec.summary}

In this work we have shown how a Floquet system acquires energy from the external drive and heats up via inter-particle interactions. Using the scattering theory of Floquet-Bloch states, we have calculated the collisional heating rates for a cold atomic gas driven by time-periodic oscillating forces. We have shown that the heating of such systems can be described by a general expression by introducing the \textit{effective collisional velocity} $v_{\rm col}$ parametrizing the density of states :
\begin{equation}
\mathcal{P} \propto \rho \sigma v_{\rm col} E_0.
\end{equation}
The velocity $v_{\rm col}$ is determined by the dominant energy in the system and is summarized in Table~\ref{tb:eff}. For fermionic systems, the collisional heating is further suppressed by Pauli blocking. In systems with lower dimensions, collisional heating is also reduced due to the modified density of states.
\renewcommand{\arraystretch}{2}
\begin{table*}[ht]
\centering
\caption[Heating Rate.]{\label{tb:eff}Effective collisional velocities $v_{\rm col}$ for various systems. The heating rates follow an unified description $\mathcal{P} \propto \rho \sigma v_{\rm col} E_0$ where $E_0$ is the strength of the drive and $\rho $ is the corresponding particle density.} 
\begin{ruledtabular}
\begin{tabular} { >{\arraybackslash}m{17mm}>{\arraybackslash}m{30mm}>{\arraybackslash}m{123mm}}
$v_{\rm col}$&Condition&\multicolumn{1}{c}{Dominating Energy Scale}\\
\hline
$\sqrt{\hbar\omega/m}$&$\hbar\omega\gg E_0, k_{\rm b}T$& \textit{Rapid-modulation regime}, where $\hbar\omega$ dominates\\
$\sqrt{k_{\rm b}T/m}$&$k_{\rm b}T \gg \hbar\omega, E_0$& \textit{Semiclassical regime}, where the thermal motion dominates\\
$\sqrt{E_0/m}$&$E_0 \gg \hbar\omega, k_{\rm b}T$& \textit{Strong-drive regime} where micromotion dominates as in, for example,  condensates and cold atomic samples.\\
$\sqrt{E_{\rm F}/m}$&$E_{\rm F} \gg \hbar\omega, k_{\rm b}T, E_0$& Fermi energy dominant as in a degenerate Fermi gas. The heating rate is further suppressed by Pauli blocking with a suppression factor, $(\hbar\omega/E_{\rm F})^2, (E_0/E_{\rm F})$, or $(k_{\rm b}T/E_{\rm F})^2$,  depending on the relation between $\hbar\omega, k_{\rm b}T, E_0$.\\
$\hbar/ml_0$&$\hbar\nu_{\perp}\gg \hbar\omega, k_{\rm b}T, E_0$& Confinement energy dominant as in systems of lower dimension.\\
\end{tabular}
\end{ruledtabular}
\end{table*}

Our calculation can also help to understand the collisional heating in other similar Floquet systems by using appropriate interparticle potentials and Floquet-Bloch states wave functions. One such system is a \textit{combined} trap for ions and neutral atoms, where ions are sympathetically cooled by atoms, limited by heating effects due to the micromotion of the ions~\cite{DeVoe2009,Chen2014,Zipkes2010,Haze2018}

In this work, we considered collisions between particles which are periodically driven by opposite forces. This is different from radio-frequency ion traps or the Time-Orbital Potential (TOP) trap~\cite{Petrich1995}, where all particles experience the same periodic force. In those cases, heating occurs due to non-adiabatic motion and the inhomogeneous strength of the drive or long-range Coulomb interactions~\cite{Wineland1998}.

A major motivation for studying Floquet heating is to assess the feasibility of preparing interesting Floquet many-body states. An essential question is whether a quantum state can be prepared before excessive heating occurs. This is often captured by a dimensionless parameter $\eta$ defined as:
\begin{equation}
\eta\sim\frac{\tau}{\tau_{\rm ev}},
\end{equation}
which characterizes the number of cycles of evolution the system can experience before the system's total energy increases by its characteristic energy $E$ due to the heating. Here $\tau_{\rm ev}$ is the timescale for the system's evolution and $\tau\sim \left(E/\mathcal{P}\right)$ is the system's lifetime. Floquet engineering of quantum states requires $\eta \gg 1$. We can use the results of this paper to estimate the parameter $\eta$ for various systems. For a thermal ensemble with $\tau_{\rm ev} \sim n_{\rm 3D} \sigma v_{\rm th}$ and $E=k_{\rm b}T$, we obtain $\eta\sim k_{\rm b}T/E_0$ in the \textit{semiclassical regime} and $\eta\sim k_{\rm b}T/E_0 \sqrt{k_{\rm b}T/\hbar\omega}$ for rapid modulation. For condensates, evolution of the system is characterized by the mean-field interaction strength $\hbar/\tau_{\rm ev}\sim U = gn_{\rm 3D} $ in 3D, which leads to $\eta\sim U^2/\mathcal{P}\sim n_{\rm 3D}/k^3_0$ or $\sim \sqrt{\hbar^2/ma^2}/\sqrt{\hbar\omega}(U/E_0)$. For Fermi gases, $\hbar/\tau_{\rm ev}\sim E_{\rm F}$, giving $\eta \sim E_{\rm F}^{1/2}/(\hbar^2\omega^2E_0)$. This illustrates the benefit of using systems with large Fermi energy. 

The purpose of this paper was a transparent treatment of heating in different regimes for a particularly simple Floquet system. Our discussion provides a starting point for more complex systems where we expect similar regimes depending on the hierarchy of the relevant energy scales.

\begin{acknowledgments}
We would like to thank useful discussions with Nigel Cooper, Erich Mueller, Alan O. Jamison, Jeongwon Lee, Furkan~\c{C}a\u{g}r{\i}~Top, Yair Margalit,  and Gediminas Juzeli\ifmmode \bar{u}\else \={u}\fi{}nas. We would like to thank Jesse Amato-Grill for critical reading of the manuscript. We acknowledge support from the NSF through the Center for Ultracold Atoms and award 1506369, from ARO-MURI Non-equilibrium Many-body Dynamics (Grant No. W911NF-14-1-0003), from AFOSR-MURI Quantum Phases of Matter (Grant No. FA9550-14-1-0035), from ONR (Grant No. N00014-17-1-2253) and a Vannevar-Bush Faculty Fellowship.
\end{acknowledgments}

\appendix
\section{Ensemble Averaged Heating Rates}
\subsection{3D Systems}\label{app.3d}
We present the detailed calculation for Eq.~(\ref{eq.heating_thermal}). Rewriting Eq.~(\ref{eq.ensemble}) with the center-of-mass and relative coordinates $\vc{K} =( \vc{k}_1 + \vc{k}_2)/2 = (\vc{k}'_1+\vc{k}'_2)/2$, and $\Gamma(\vc{k}_1,\vc{k}_2\rightarrow\vc{k}'_1,\vc{k}'_2) = \Gamma(\vc{k}\rightarrow\vc{k'})$ with $\vc{k} = (\vc{k}_1-\vc{k}_2),\vc{k'}=(\vc{k}'_1-\vc{k}'_2)$ one obtains:
\begin{equation}
\mathcal{P} = \int \frac{{\rm d}^3\vc{K}}{(2\pi)^3}\frac{{\rm d}^3\vc{k}}{(2\pi)^3}\frac{{\rm d}^3\vc{k'}}{(2\pi)^3}f(\frac{\vc{k}}{2}+\vc{K})f(-\frac{\vc{k}}{2}+\vc{K})\mathcal{P}(\vc{k}\rightarrow\vc{k}')
\end{equation}
with the Boltzmann distribution function. 
$$
f(\vc{k}) =N \left(\frac{\hbar^2}{2\pi mk_{\rm b}T}\right)^{3/2}\exp{\left(-\frac{\hbar^2\vc{k}\cdot\vc{k}}{2m k_{\rm b}T}\right)}
$$

As discussed in Sec.~\ref{sec.regimes}, we consider only the processes where $n = \pm 1$. The coupling matrix elements $M_{\rm n}$, to the lowest orders in $\alpha_{k}$, are shown in Table II.
\begin{table}[ht]
\centering
\caption[Heating Rate.]{Approximate coupling matrix elements $M_{\rm n}$ for various Floquet collision processes.} 
\begin{ruledtabular}
\begin{tabular} { >{\center\arraybackslash}m{4mm}>{\center\arraybackslash}m{17mm}>{\center\arraybackslash}m{17mm}>{\center\arraybackslash}m{21mm}>{\center\arraybackslash}m{21mm}}
$M_0/g$&$M_{1}/g$&$M_{-1}/g$&$M_{2}/g$&$M_{-2}/g$\\
\hline
1&$\frac{\hbar k_0(k_z-k'_z)}{2m\omega}$&$-\frac{\hbar k_0(k_z-k'_z)}{2m\omega}$&$\frac{1}{2}\left[\frac{\hbar k_0(k_z-k'_z)}{2m\omega}\right]^2$&$\frac{1}{2}\left[\frac{\hbar k_0(k_z-k'_z)}{2m\omega}\right]^2$\\
\end{tabular}
\end{ruledtabular}
\end{table}

With the single-sideband assumption, we obtain the analytic expression of Eq.~(\ref{eq.general_heating}) by explicitly calculating $\gamma^2(\vc{k},\pm 1)$ :
\begin{equation}
\begin{split}
\mathcal{P}_{\vc{k}} 
& =4 \frac{\sigma}{V}\sum_{n = \pm 1}\sqrt{\frac{E_{\vc{k}}+n\hbar\omega}{m}}\left(\frac{\hbar^2k_z^2}{\mu\hbar\omega}+\frac{\hbar^2|\vc{k}|^2}{3\mu\hbar\omega}+\frac{2n}{3}\right)n E_0.\\
\end{split}
\label{eq:transition_n1}
\end{equation}
with $n=\pm 1$ for the \textit{semiclassical regime}, and $n = 1$ for the \textit{rapid-modulation regime}. The total heating rate $\mathcal{P}$ can be readily obtained as
\begin{equation}
\begin{split}
\mathcal{P} = &N^2\left(\frac{4\pi \hbar^2}{mk_{\rm b} T}\right)^{3/2}\int \frac{{\rm d}^3\vc{k}}{(2\pi)^3} e^{-\frac{\hbar^2\vc{k}\cdot\vc{k}}{4mk_{\rm b}T}}\mathcal{P}_{\vc{k}} ,\\
\end{split}
\end{equation}
which gives Eq.~(\ref{eq.heating_bec}) and Eq.~(\ref{eq.heating_thermal}) at the corresponding limits.
\subsection{Quasi-2D Systems}\label{app.2d}
We start with the Floquet-Bloch states wave function for quasi-2D system written as Eq.~(\ref{eq.2d_wf}). The scattering strength, defined as $$g = \int {\rm d}\vc{r}\;\phi_i(\vc{r})V(\vc{r})\phi_f(\vc{r}),$$ has the form~\cite{Petrov2000}:
$$g_{\rm 2D} = \frac{2\sqrt{2\pi}\hbar^2}{m}\frac{a}{l_0}$$
in quasi-2D. The results obtained for 3D cases can be readily extended to quasi-2D by replacing $g$ with $g_{\rm 2D}$ and using the corresponding 2D density. We therefore obtain the coupling matrix element between two quasi-2D Floquet-Bloch states
\begin{equation}
M_{n}(\vc{k}\rightarrow \vc{k'}) = g_{\rm 2D}J_{n}(\alpha_k -\alpha_{k'}),
\end{equation}
and obtain
\begin{equation}
\begin{split}
&\gamma^2_{\rm 2D}(\vc{k},n) = \\
&\frac{1}{2g_{\rm 2D}^2}\int_0^{\pi} {\rm d}\theta\;|M_n(\vc{k}\rightarrow\sqrt{|\vc{k}|^2+2\mu n\omega/\hbar}\cos\theta)|^2.
\end{split}
\end{equation}
Following similar procedures as in the 3D calculation, and using the 2D density of states lead to
\begin{equation}
\begin{split}
\mathcal{P}^{\rm 2D}_{\vc{k}} 
& = \frac{\pi \mu g^2_{\rm 2D}}{4\hbar^3 A}\sum_{n = \pm 1}\left(\frac{2\hbar^2k_z^2}{\mu\hbar\omega}+\frac{\hbar^2|\vc{k}|^2}{\mu\hbar\omega}+2n\right)n E_0.\\
\end{split}
\end{equation}
Together with the 2D Boltzmann distribution
$
f(\vc{k}) =N \left(\frac{2\pi\hbar^2}{m k_{\rm b}T}\right)\exp{\left(-\frac{\hbar^2\vc{k}\cdot\vc{k}}{2m k_{\rm b}T}\right)},
$
we obtain the heating rate of the ensemble 
\begin{equation}
\begin{split}
\mathcal{P}_{2\rm D} =& \int \frac{{\rm d}^2\vc{K}}{(2\pi)^2}\frac{{\rm d}^2\vc{k}}{(2\pi)^2}\frac{{\rm d}^2\vc{k'}}{(2\pi)^2} f(\frac{\vc{k}}{2}+\vc{K})f(-\frac{\vc{k}}{2}+\vc{K})\mathcal{P}(\vc{k}\rightarrow\vc{k}')\\
=&2N^2\left(\frac{2\pi\hbar^2}{mk_{\rm b}T}\right)\int \frac{{\rm d}^2\vc{k}}{(2\pi)^2} e^{-\frac{\hbar^2\vc{k}\cdot\vc{k}}{4mk_{\rm b}T}}\mathcal{P}^{\rm 2D}_{\vc{k}},
\end{split}
\end{equation}
which gives Eqs.~(\ref{eq.2dfast}) and~(\ref{eq.2dsemi}).
\subsection{Quasi-1D Systems}\label{app.1d}
For a quasi-1D system, we use the 1D wave function, a scattering strength $g_{\rm 1D}$, and $M_{\rm n}$
\begin{equation}
\begin{split}
&\Psi(x, y, z,t) = \phi_{\perp}(x, y)\frac{1}{\sqrt{L}}\exp{\left[ik(t)z -i\Phi(t)\right]},\\
&g_{\rm 1D} = \frac{\hbar^2a}{ml^2_0},\\
&M_{n}(k\rightarrow k') = g_{\rm 1D}J_{n}(\alpha_k -\alpha_{k'}),\\
&D_{\rm 1D}(E) = \frac{1}{2\pi\hbar}\sqrt{\frac{\mu}{2E}},
\end{split}
\label{eq.1d_wf}
\end{equation}
to obtain
\begin{equation}
\begin{split}
\gamma^2_{\rm 1D}(k,n) =& \\
\frac{1}{2g_{\rm 1D}^2}&\bigg[|M_n(k\rightarrow\sqrt{k^2+2\mu n\omega/\hbar})|^2\\
&+|M_n(k\rightarrow-\sqrt{k^2+2\mu n\omega/\hbar})|^2\bigg],
\end{split}
\end{equation}
yielding
\begin{equation}
\begin{split}
\mathcal{P}^{\rm 1D}_{k} 
& = \frac{8g^2_{\rm 1D}}{\hbar^2\omega L}\sum_{n }\frac{k^2 + n\mu\hbar\omega/\hbar^2}{\sqrt{k^2+2\mu n\hbar\omega/\hbar^2}}n E_0.\\
\end{split}
\end{equation}
With the 1D Boltzmann distribution
$
f(k) =N\left(\frac{2\pi\hbar^2}{mk_{\rm b}T}\right)^{1/2}\exp{\left(-\frac{\hbar^2k^2}{m k_{\rm b}T}\right)},
$ we calculate the heating rate of the ensemble 
\begin{equation}
\begin{split}
\mathcal{P}_{1\rm D} =& \int \frac{{\rm d}K}{2\pi}\frac{{\rm d}k}{2\pi}\frac{{\rm d} k'}{2\pi} f(\frac{k}{2}+K)f(-\frac{k}{2}+K)\mathcal{P}(k\rightarrow k')\\
=&N^2\left(\frac{\pi\hbar^2}{mk_{\rm b}T}\right)^{1/2}\int \frac{{\rm d}k}{2\pi} e^{-\frac{\hbar^2k^2}{4mk_{\rm b}T}}\mathcal{P}^{\rm 1D}_{\rm k},
\end{split}
\end{equation}
which results in Eq.~(\ref{eq.1d}). 

We add the note that the scaling $\sim 1/\sqrt{k_{\rm b}T}$ in the regime $\hbar\omega \ll k_{\rm b}T$ \textit{cannot} be directly obtained in the same way as for other dimensions. In 1D, terms in zeroth, first, and second order in $\hbar\omega/E_{k}$ cancel between the $n=\pm 1$ processes if both processes are allowed for two particles with $k^2 >2\mu n \hbar\omega/h^2$
\begin{equation}
\begin{split}
&\sum_{n = \pm 1 }\frac{k^2 + n\mu\hbar\omega/\hbar^2}{\sqrt{k^2+2\mu n\hbar\omega/\hbar^2}}n \\ &= -\left(\frac{\hbar\omega}{E_k}\right)^3\left(\frac{\hbar^2}{mE_k}\right)^2+\mathcal{O}\left[\left(\frac{\hbar\omega}{E_k}\right)^4\right]. \\
\end{split}
\end{equation}
Therefore, even for $k_{\rm b}T \gg \hbar\omega$, the leading contribution comes from the regime $k^2 < 2\mu n \hbar\omega/h^2 $, where only the $n=+1$ process is allowed. We obtain
\begin{equation}
\begin{split}
\mathcal{P}_{1\rm D} =&N^2\left(\frac{\pi\hbar^2}{mk_{\rm b}T}\right)^{1/2}\int \frac{{\rm d}k}{2\pi} e^{-\frac{\hbar^2k^2}{4mk_{\rm b}T}}\mathcal{P}^{\rm 1D}_{\rm k}\\
\approx &N^2 \left(\frac{\hbar}{m}\right)^3\frac{\sigma}{l_0^4}\frac{E_0}{\pi^2\omega L}\sqrt{\frac{m\pi}{k_{\rm b}T}}\\
&\times  \int_{-\sqrt{2\mu \hbar\omega/\hbar^2}}^{\sqrt{2\mu \hbar\omega/\hbar^2}} {\rm d}k e^{-\frac{\hbar^2k^2}{4mk_{\rm b}T}}\frac{k^2 + \mu\hbar\omega/\hbar^2}{\sqrt{k^2+2\mu \hbar\omega/\hbar^2}}\\\\
\sim &N^2 \frac{\sqrt{2}}{\pi^2L}\left(\frac{\hbar}{m}\right)^2\sqrt{\frac{m\pi}{k_{\rm b}T}}\frac{\sigma}{l_0^4}E_0 \;\;\;\;(\hbar\omega\ll k_{\rm b}T).
\end{split}
\end{equation}
The obtained scaling is consistent with the numerical results.

 \section{Derivation of the Pauli Blocking Factor}\label{app.fermi}
\subsection{Zero-Temperature Fermi Gases}
In this part we present the calculation for Eq.~(\ref{eq.fermi_3D}). The interspin collision rate of a two-component Fermi mixture is written as
\begin{equation}
\begin{split}
\mathcal{P}^{\rm F}_{\rm ens}= &\int \frac{{\rm d^3}\vc{k}_1}{(2\pi)^3} \frac{{\rm d^3}\vc{k}_2}{(2\pi)^3}\frac{{\rm d^3}\vc{k_1'}}{(2\pi)^3} \frac{{\rm d^3}\vc{k_2'}}{(2\pi)^3}\delta_{\vc{k}_1+\vc{k}_2,\vc{k}'_1+\vc{k}'_2}\\
&\times f_{\downarrow}(\vc{k}_1)f_{\uparrow}(\vc{k}_2)(1-f_{\downarrow}(\vc{k}'_1))(1-f_{\uparrow}(\vc{k}'_2))\\
&\times\mathcal{P}(\vc{k}_1,\vc{k}_2\rightarrow\vc{k}'_1,\vc{k}'_2).\\
\end{split}
\label{eq:fermi_collision_rate}
\end{equation}
We denote $\epsilon_i = \hbar^2|\vc{k}_i|^2/2m$, $\epsilon'_i = \hbar^2|\vc{k}'_i|^2/2m$, and rewrite the integral in spherical coordinates with $\vc{k}_i = k_i(\sin{\theta_i}\cos{\phi_i},\sin{\theta_i}\sin{\phi_i},\cos{\theta_i}),\vc{k}'_i = k'_i(\sin{\theta'_i}\cos{\phi'_i},\sin{\theta'_i}\sin{\phi'_i},\cos{\theta'_i})$.

Momentum conservation $\vc{k}_1+\vc{k}_2 = \vc{k}'_1+\vc{k}'_2$ gives
\begin{equation}
\begin{split}
&\mathcal{P}(\vc{k}_1,\vc{k}_2\rightarrow\vc{k}'_1,\vc{k}_1+\vc{k}_2-\vc{k}'_1) \\
&= C_{\Gamma}\left[\frac{(\vc{k}_1-\vc{k}_2)_z-(\vc{k}'_1-\vc{k}_2')_z}{4}\right]^2\delta (\epsilon_1+\epsilon_2 +\hbar\omega-\epsilon_1'-\epsilon_2')\\
&= C_{\Gamma}\left[\frac{k_1\cos{\theta_1}-k'_1\cos{\theta_1'}}{2}\right]^2\delta (\epsilon_1+\epsilon_2 +\hbar\omega-\epsilon_1'-\epsilon_2'),
\end{split}
\end{equation}
where $C_{\Gamma}=\frac{2\pi}{\hbar}\frac{g^2\hbar^3k_0^2}{\mu^2\omega}$ is a constant, and $\epsilon_2' = \frac{\hbar^2}{2m}|\vc{k}_1+\vc{k}_2-\vc{k}'_1|^2$. The heating rate Eq.~(\ref{eq:fermi_collision_rate}) now has the form
\begin{equation}
\begin{split}
\mathcal{P}^{\rm F}_{\rm ens}= &\int \frac{{\rm d^3}\vc{k}_1}{(2\pi)^3} \frac{{\rm d^3}\vc{k}_2}{(2\pi)^3}\frac{{\rm d^3}\vc{k_1'}}{(2\pi)^3}\delta (\epsilon_1+\epsilon_2 +\hbar\omega-\epsilon_1'-\epsilon_2')\\
\times & f_{\downarrow}(\vc{k}_1)f_{\uparrow}(\vc{k}_2)(1-f_{\downarrow}(\vc{k}'_1))(1-f_{\uparrow}(\vc{k}_1+\vc{k}_2-\vc{k}'_1))\\
\times & C_{\Gamma}\left[\frac{k_1\cos{\theta_1}-k'_1\cos{\theta_1'}}{2}\right]^2\\
= &C_{\Gamma}\left[\frac{(2m)^{3/2}}{2\hbar^3}\right]^3\int \frac{\sqrt{\epsilon_1\epsilon_2\epsilon'_1}}{(2\pi)^9}{\rm d}\epsilon_1 {\rm d}\epsilon_2{\rm d}\epsilon'_1\;{\rm d} \Omega_1\;{\rm d} \Omega_2\;{\rm d} \Omega'_1\\
\times & f_{\downarrow}(\epsilon_1)f_{\uparrow}(\epsilon_2)(1-f_{\downarrow}(\epsilon'_1))(1-f_{\uparrow}(\epsilon_1+\epsilon_2+\hbar\omega -\epsilon'_1))\\
\times & \left[\frac{k_1\cos{\theta_1}-k'_1\cos{\theta_1'}}{2}\right]^2\\
\times &\delta (\epsilon_1+\hbar\omega-\epsilon'_1- \frac{\hbar^2q^2}{2m}+\frac{\hbar^2}{m}\vc{q}\cdot\vc{k}_2),
\end{split}
\end{equation}
where $\vc{q} = \vc{k}_1-\vc{k}'_1$ and $q = |\vc{q}|$. Integrating over the solid angle ${\rm d}\Omega_2  = \sin{\theta_2}{\rm d}\theta_2{\rm d}\phi_2$ gives
\begin{equation}
\begin{split}
&\mathcal{P}^{\rm F}_{\rm ens}=(2\pi)C_{\Gamma}\left[\frac{(2m)^{3/2}}{2\hbar^3}\right]^3  \int \frac{\sqrt{\epsilon_1\epsilon_2\epsilon_1'}}{(2\pi)^9} {\rm d}\epsilon_1{\rm d}\epsilon_2{\rm d}\epsilon'_1\;{\rm d} \Omega_1\;{\rm d} \Omega'_1\\
&\times f_{\downarrow}(\epsilon_1)f_{\uparrow}(\epsilon_2)(1-f_{\downarrow}(\epsilon'_1))(1-f_{\uparrow}(\epsilon_1+\epsilon_2+\hbar\omega -\epsilon'_1))\\
&\times \left[\frac{k_1\cos{\theta_1}-k'_1\cos{\theta_1'}}{2}\right]^2\\
&\times\frac{m}{\hbar^2q\sqrt{\epsilon_2}} \int_{-1}^1{\rm d}x\;\delta (x-\frac{-\epsilon_1-\hbar\omega+\epsilon'_1+\frac{\hbar^2q^2}{2m}}{\frac{\hbar^2}{m}q\sqrt{\epsilon_2}}),\\
\end{split}
\end{equation}
with $x = \cos\theta_2$. We subsequently integrate over $\Omega_1$ and $\Omega'_1$. When $\vc{k}_1,\vc{k}'_1$ vary over their respective solid angles $\Omega_1, \Omega_1'$ with fixed lengths $k_1 = \sqrt{\epsilon_1},k_1' = \sqrt{\epsilon'_1}$, the differential vector $\vc{q}$ also varies over the entire solid angle $\Omega_q $ with the length varying from $\sqrt{\epsilon_1'}-\sqrt{\epsilon_1}$ to $\sqrt{\epsilon_1'}+\sqrt{\epsilon_1}$. Therefore, we have the equivalence
$$
{\rm d}\Omega_1{\rm d}\Omega_2 = C_qq^2{\rm d}q\sin{\theta_q}{\rm d}\theta_q{\rm d}\phi_q {\rm d}\phi_l,
$$
where $C_q = 6/(\sqrt{\epsilon_1'}+\sqrt{\epsilon_1})^3$ is the normalization factor, and $\phi_l$ is the angle between the planes expanded by $\{\vc{k}_1,\vc{k}_1'\}$ and $\{\vc{q},\vc{e}_{z}\}$. The heating rate follows as
\begin{equation}
\begin{split}
&\mathcal{P}^{\rm F}_{\rm ens}=(2\pi)^3 C_{\Gamma}\left[\frac{(2m)^{3/2}}{2\hbar^3}\right]^3\int \frac{C_q \sqrt{\epsilon_1\epsilon_2\epsilon_1'}}{(2\pi)^9} {\rm d}\epsilon_1{\rm d}\epsilon_2{\rm d}\epsilon'_1 \\
&\times q^2{\rm d}q\sin{\theta_q}{\rm d}\theta_q\\
&\times f_{\downarrow}(\epsilon_1)f_{\uparrow}(\epsilon_2)(1-f_{\downarrow}(\epsilon'_1))(1-f_{\uparrow}(\epsilon_1+\epsilon_2+\hbar\omega -\epsilon'_1))\\
&\times\frac{m(q\cos{\theta_q})^2}{4\hbar^2q\sqrt{\epsilon_2}} \int\displaylimits_{-1}^1{\rm d}x\;\delta (x-\frac{-\epsilon_1-\hbar\omega+\epsilon'_1+\frac{\hbar^2q^2}{2m}}{\hbar^2 q\sqrt{\epsilon_2}/m})\\
&=C_{\Gamma}\left[\frac{(2m)^{3/2}}{2\hbar^3}\right]^3 \int \frac{C_q \sqrt{\epsilon_1\epsilon_2\epsilon_1'}}{(2\pi)^6} {\rm d}\epsilon_1{\rm d}\epsilon_2{\rm d}\epsilon'_1\\
&\times f_{\downarrow}(\epsilon_1)f_{\uparrow}(\epsilon_2)(1-f_{\downarrow}(\epsilon'_1))(1-f_{\uparrow}(\epsilon_1+\epsilon_2+\hbar\omega -\epsilon'_1))\\
&\times \int\displaylimits _{\sqrt{\epsilon'_1}-\sqrt{\epsilon_1}}^{\sqrt{\epsilon'_1}+\sqrt{\epsilon_1}}\frac{mq^3{\rm d}q}{6\hbar^2\sqrt{\epsilon_2}} \int\displaylimits_{-1}^1{\rm d}x\;\delta (x-\frac{-\epsilon_1-\hbar\omega+\epsilon'_1+\frac{\hbar^2q^2}{2m}}{\hbar^2q\sqrt{\epsilon_2}/m})\\
&= C_{\Gamma}\left[\frac{(2m)^{3/2}}{2\hbar^3}\right]^3 \int \frac{C_q \sqrt{\epsilon_1\epsilon_2\epsilon_1'}}{(2\pi)^6} {\rm d}\epsilon_1{\rm d}\epsilon_2{\rm d}\epsilon'_1 \\
&\times f_{\downarrow}(\epsilon_1)f_{\uparrow}(\epsilon_2)(1-f_{\downarrow}(\epsilon'_1))(1-f_{\uparrow}(\epsilon_1+\epsilon_2+\hbar\omega -\epsilon'_1))\\
&\times\bigg[ \int_{\sqrt{\epsilon'_1}-\sqrt{\epsilon_1}}^{\sqrt{\epsilon'_1}+\sqrt{\epsilon_1}}\frac{mq^3{\rm d}q }{6\hbar^2\sqrt{\epsilon_2}} \\
&\theta (\sqrt{2\epsilon'_2-\epsilon_2}+\sqrt{\epsilon_2}-q)\theta(q-(\sqrt{2\epsilon'_2-\epsilon_2}-\sqrt{\epsilon_2})) \bigg].\\\
\end{split}
\label{eq.fermi_heating_2}
\end{equation}
Note that so far the only approximation adopted is the single-sideband approximation, and that the first sideband is weak, implying $J_1(x) \approx x/2$.

When $\hbar\omega < E_{\rm F}$, we have $\sqrt{2\epsilon'_2-\epsilon_2}+\sqrt{\epsilon_2}\approx \sqrt{\epsilon'_1}+\sqrt{\epsilon_1}\approx 2\sqrt{\hbar^2/(2m)}k_{\rm F}$, and $\sqrt{2\epsilon'_2-\epsilon_2}-\sqrt{\epsilon_2}\approx \sqrt{\epsilon'_1}-\sqrt{\epsilon_1}\approx \hbar\omega/(2\sqrt{E_{\rm F}})$. The integral over $q$ in Eq.~\ref{eq.fermi_heating_2} now gives  
\begin{equation}
\begin{split}
&\mathcal{P}^{\rm F}_{\rm ens}\approx C_{\Gamma}\left[\frac{(2m)^{3/2}}{2\hbar^3}\right]^3 \int \frac{C_q\sqrt{\epsilon_1\epsilon_2\epsilon'_1}}{(2\pi)^6}{\rm d}\epsilon_1 {\rm d}\epsilon_2{\rm d}\epsilon'_1\\
&\times f_{\downarrow}(\epsilon_1)f_{\uparrow}(\epsilon_2)(1-f_{\downarrow}(\epsilon'_1))(1-f_{\uparrow}(\epsilon_1+\epsilon_2+\hbar\omega -\epsilon'_1))\\
&\times \frac{8m k_{\rm F}^4}{3\hbar^2\sqrt{\epsilon_2}}\left(\frac{\hbar^2}{2m}\right)^2\\
&\approx C_{\Gamma}\left[\frac{(2m)^{3/2}}{2\hbar^3}\right]^3 E^{3/2}_{\rm F}\frac{(\hbar\omega)^3}{8(2\pi)^6}\frac{8m}{3\hbar^2}.\\
\end{split}
\label{eq.final}
\end{equation}
Here we adopt the approximation
\begin{equation}
\begin{split}
\int &C_q\sqrt{\epsilon_1\epsilon_2\epsilon'_1}{\rm d}\epsilon_1 {\rm d}\epsilon_2{\rm d}\epsilon'_1\frac{1}{\sqrt{\epsilon_2}}\\
&\times f_{\downarrow}(\epsilon_1)f_{\uparrow}(\epsilon_2)(1-f_{\downarrow}(\epsilon'_1))(1-f_{\uparrow}(\epsilon_1+\epsilon_2+\hbar\omega -\epsilon'_1))\\
\approx & E^{-1/2}_{\rm F}\frac{(\hbar\omega)^3}{ 8}.\\
\end{split}
\end{equation}

With the definition of the Fermi energy $(2mE_{\rm F})^{3/2} = 3n_{\rm 3D}\hbar^3\pi^2$, we eventually obtain
\begin{equation}
\mathcal{P}^{\rm F}_{\rm ens}\approx \frac{\pi}{ \sqrt{2}}N n_{\rm 3D}  \sigma\left(\frac{\hbar\omega}{E_{\rm F}}\right)^2\sqrt{\frac{E_{\rm F}}{m}}  E_0,
\end{equation}
which shows explicitly the Fermi suppression factor $\left(\hbar\omega/E_{\rm F}\right)^2$.

The calculations above can be extended to the regime of small $\hbar\omega$ where multi-quanta transfer processes are relevant.

\subsection{Finite Temperature Fermi Gases}
In this section we present the calculation for the Pauli blocking factor $(T/T_{\rm F})^2$ when $T/T_{\rm F} \ll 1$. We further assume that $\hbar\omega \ll k_{\rm b}T$. The non-zero temperature case is different from the zero temperature Fermi gas mainly in two aspects: first, as discussed briefly in the main text, the active Fermi shell formed by the accessible states has a thickness of $k_{\rm b} T$ instead of $\hbar\omega$. Second, energy quanta \textit{emission} processes are now allowed. In the lowest order approximation, the heating rate can be calculated with only $n\pm 1$ processes and leads to
\begin{equation}
\begin{split}
\int &C_{q}\sqrt{\epsilon_1\epsilon_2\epsilon'_1}{\rm d}\epsilon_1 {\rm d}\epsilon_2{\rm d}\epsilon'_1\frac{1}{\sqrt{\epsilon_2}}\\
&\times \bigg[f_{\downarrow}(\epsilon_1)f_{\uparrow}(\epsilon_2)(1-f_{\downarrow}(\epsilon'_1))(1-f_{\uparrow}(\epsilon_1+\epsilon_2+\hbar\omega -\epsilon'_1))\\
&-f_{\downarrow}(\epsilon_1)f_{\uparrow}(\epsilon_2)(1-f_{\downarrow}(\epsilon'_1))(1-f_{\uparrow}(\epsilon_1+\epsilon_2-\hbar\omega -\epsilon'_1))\bigg]\\
=&\int C_{q}\sqrt{\epsilon_1\epsilon_2\epsilon'_1}{\rm d}\epsilon_1 {\rm d}\epsilon_2{\rm d}\epsilon'_1\frac{1}{\sqrt{\epsilon_2}}\\
&\times f_{\downarrow}(\epsilon_1)f_{\uparrow}(\epsilon_2)(1-f_{\downarrow}(\epsilon'_1))\left(-\frac{\partial f(\epsilon)}{\partial\epsilon}\hbar\omega\right)\bigg\rvert_{\epsilon = \epsilon_1+\epsilon_2-\epsilon'_1}\\
\sim & E^{-1/2}_{\rm F}\left(k_{\rm b}T\right)^2\hbar\omega.\\
\end{split}
\end{equation}
Together with Eq.~(\ref{eq.final}), the Fermi suppression factor is readily recognized to be $(k_{\rm b}T/E_{\rm F})^2$.

\newpage
%

\end{document}